# Neutrino and Anti-neutrino Cross Sections at MiniBooNE

Ranjan Dharmapalan for the MiniBooNE Collabration

*The University of Alabama*
*Department of Physics and Astronomy, Tuscaloosa, AL-35487, U.S.A.*

**Abstract.** The MiniBooNE experiment has reported a number of high statistics neutrino and anti-neutrino cross sections -among which are the charged current quasi-elastic (CCQE) and neutral current elastic (NCE) neutrino scattering on mineral oil ($CH_2$). Recently a study of the neutrino contamination of the anti-neutrino beam has concluded and the analysis of the anti-neutrino CCQE and NCE scattering is ongoing.

## MOTIVATION

Understanding the physics of neutrino scattering is important for neutrino oscillation experiments and also nuclear physics.

Charged current quasi-elastic (CCQE) interaction is the event signature for oscillation experiments such as: MiniBooNE, T2K, NoνA, etc. Especially in the era of precision experiments there is a need to reduce the cross section error. There is also a dearth in cross section information particularly in the sub-GeV range.

Neutrinos also serve as a unique nuclear probe. CCQE and neutral current elastic (NCE) interactions are sensitive to the axial mass ($M_A$) of the nucleon. NCE interactions are further sensitive to $\Delta s$ -the strange quark component of the nucleus.

## NEUTRINO MODE RESULTS

In the neutrino mode a high statistic sample of charged-current muon neutrino scattering events were collected and analyzed [1] to extract the first measurement of the double differential cross section ($d^2\sigma/dT_\mu dcos\theta_\mu$)[Fig.1] for CCQE scattering on carbon. This result featured minimal model dependence and provided the most complete information on this process to date. With the assumption of CCQE scattering , the absolute cross section as a function of neutrino energy $\sigma(E_\nu)$[Fig.2] and the single differential cross section ($d\sigma/dQ^2$)[Fig.3] were extracted to facilitate comparison with previous measurements.

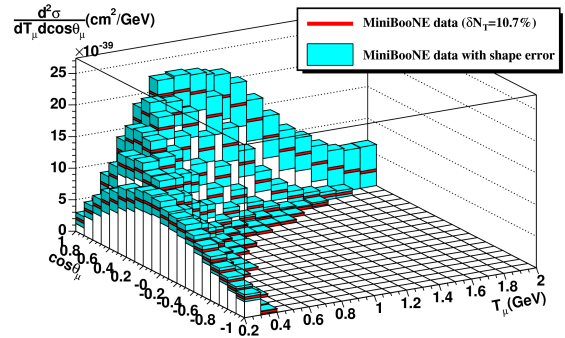

**FIGURE 1.** Flux-integrated double differential cross section per target neutron for the $\nu_\mu$ CCQE process. The dark bars indicate the measured values and the surrounding lighter bands show the shape error.

The reported cross section is significantly larger ($\approx$ 30% at the flux average energy) than what is commonly assumed for this process assuming a relativistic Fermi Gas model (RFG) and the world-average value for the axial mass, $M_A$ = 1.03 GeV [2] In addition, the $Q^2_{QE}$ distribution of this data shows a significant excess of events over this expectation at higher $Q^2_{QE}$ even if the data is normalized to the prediction over all $Q^2_{QE}$ . This leads to an extracted axial mass from a "shape-only" fit of the $Q^2_{QE}$ distribution of $M_A^{eff}$= 1.35±0.17 GeV, significantly higher than the historical world-average value.

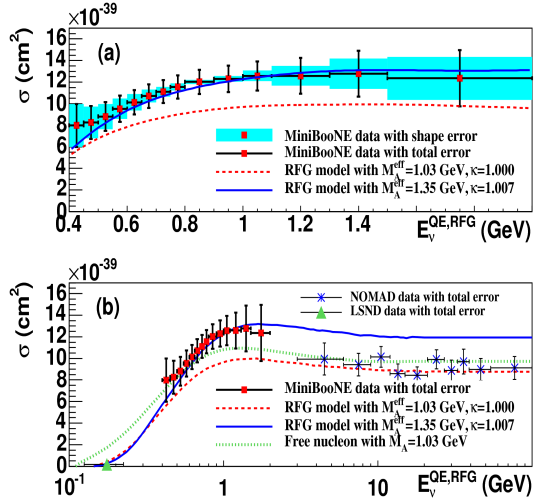

**FIGURE 2.** Flux-unfolded MiniBooNE $\nu_\mu$ CCQE cross section per neutron as a function of neutrino energy. In (a), shape errors are shown as shaded boxes along with the total errors as bars. In (b), a larger energy range is shown along with results from the LSND [3] and NOMAD [4] experiments. Also shown are predictions from the NUANCE[1] simulation for an RFG model with two different parameter variations and for scattering from free nucleons with the world average $M_A$ value.

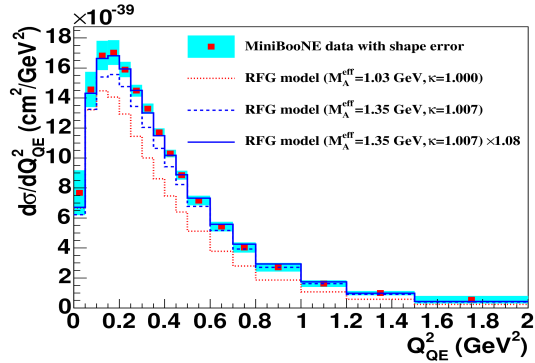

**FIGURE 3.** Flux-integrated single differential cross section per target neutron for the $\nu_\mu$ CCQE process. The measured values are shown as points with the shape error as shaded bars. Calculations from the nuance RFG model with different assumptions for the model parameters are shown as histograms.

For the NCE analysis MiniBooNE reported [6] a measurement of the flux-averaged neutral current elastic differential cross section for neutrinos scattering on mineral oil ($CH_2$) as a function of four-momentum transferred squared[Fig.4]. It was obtained by measuring the kinematics of recoiling nucleons with kinetic energy greater than 50 MeV which are readily detected in MiniBooNE. This differential cross section distribution was fit with fixed nucleon form factors apart from an axial mass, $M_A$, that provides a best fit for $M_A = 1.39\pm0.11$ GeV [Fig. 5] Additionally, single protons with kinetic energies above 350 MeV (Cerenkov threshold for protons in mineral oil) could be distinguished from neutrons and multiple nucleon events. Using this marker, the strange quark contribution to the neutral current axial vector form factor at $Q^2 = 0$, $\Delta s$, was found to be $\Delta s = 0.08 \pm 0.26$ [Fig.6].

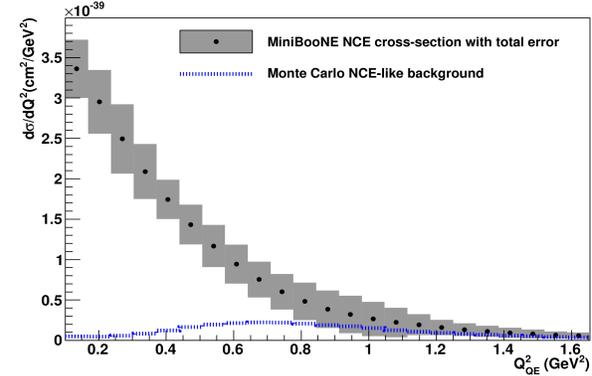

**FIGURE 4.** The MiniBooNE NCE ($\nu N \to \nu N$) flux-averaged differential cross section on $CH_2$ as a function of , $Q^2_{QE}= 2m_N \Sigma_i T_i$ where we sum the true kinetic energies of all final state nucleons produced in the NCE interaction. The blue dotted line is the predicted spectrum of NCE-like background which has been subtracted out from the total NCE-like differential cross section.

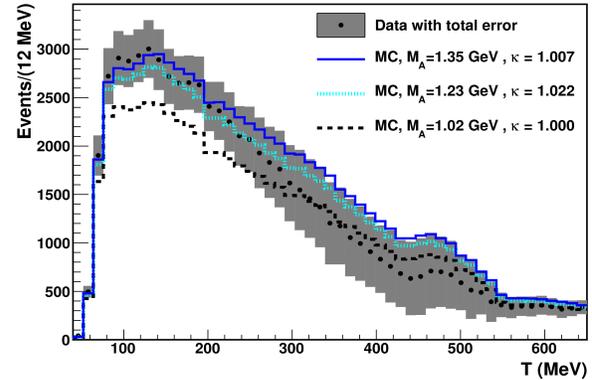

**FIGURE 5.** $\chi^2$-tests performed on the NCE reconstructed nucleon kinetic energy distribution for MC with different $M_A$ values of 1.35, 1.23, and 1.02 GeV. The $\chi^2$ values are 27.1, 29.2 and 41.3 for 49 degrees of freedom (DOF), respectively. The distributions are absolutely normalized.

---

[1] The MiniBooNE experiment employs the NUANCE v3 event generator [5] to estimate neutrino interaction rates in the $CH_2$ target medium.

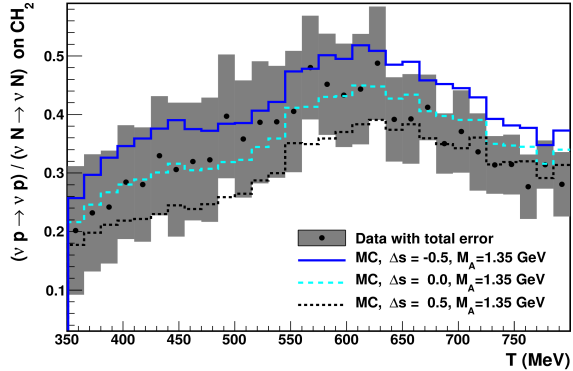

**FIGURE 6.** The ratio of νp → νp/νN → νN as a function of the reconstructed energy for data and MC with Δs values as labeled

## ANTI-NEUTRINO MODE

Before moving on to anti-neutrino cross section measurements it is important to calibrate the neutrino component of the anti-neutrino beam -the Wrong Sign(WS) contamination. The WS contamination is a greater concern in the anti-neutrino mode than in the neutrino mode due to effects from both flux and cross section: the leading particle effect at the target preferentially produces about twice as many $\pi^+$ as $\pi^-$, and the neutrino cross section is about three times higher than the anti-neutrino cross section in the MiniBooNE energy range (∼ 1 GeV).

The study of WS contamination in MiniBooNE has recently concluded.

In the anti-neutrino mode MiniBooNE plans to report both, the CCQE and NCE cross section. The sample size in both channels, represent the largest to date.

## Other cross section studies at MiniBooNE

MiniBooNE has reported other cross section studies such as : neutral current $\pi^0$ cross sections on mineral oil [7], charged current $\pi^+$/quasi-elastic cross section ratio on mineral oil [8], charged current $\pi^0$ production cross sections on mineral oil[9] and charged current $\pi^+$ production cross sections on mineral oil [10].

## ACKNOWLEDGMENTS

We are grateful to the organizers of NuFact 2010 for the opportunity to present a poster.